\documentclass[aps,prd,superscriptaddress, nofootinbib,12pt]{revtex4-2}
\usepackage{graphicx,amsmath,amssymb}
\usepackage{bigints}
\usepackage{bm}
\usepackage{ulem}
\usepackage{hyphenat}
\usepackage{setspace}\usepackage[bottom]{footmisc}
\usepackage{hyperref}
\hypersetup{colorlinks=true, linkcolor = [rgb]{0,0.08,0.45}, citecolor = [rgb]{0,0.08,0.45}, urlcolor = [rgb]{0,0.08,0.45}}

\newcommand{\half}{{\textstyle{\frac{1}{2}}}}

\newcommand{\beq}{\begin{equation}}
\newcommand{\enq}{\end{equation}}
\newcommand{\beqa}{\begin{eqnarray}}
\newcommand{\beqast}{\begin{eqnarray*}}
\newcommand{\enqa}{\end{eqnarray}}
\newcommand{\enqast}{\end{eqnarray*}}
\newcommand{\nn}{\nonumber}
\newcommand{\req}[1]{(\ref{#1})}
\newcommand{\lb}{\label}

\newcommand{\pa}{\partial}

\newcommand{\bec}{\begin{center}}
\newcommand{\enc}{\end{center}}
\newcommand{\beqo}{\begin{quote}}
\newcommand{\enqo}{\end{quote}}

\newcommand{\mbf}[1]{\mathbf{#1}}

\newcommand{\cN}{{\cal N}}

\newcommand{\cS}{{\cal S}}

\newcommand{\al}{\alpha}

\newcommand{\ga}{\gamma}

\newcommand{\ze}{\zeta}

\newcommand{\ka}{\kappa}
\newcommand{\la}{\lambda}
\newcommand{\rh}{\rho}
\newcommand{\si}{\sigma}

\newcommand{\ta}{\tau}

\newcommand{\ph}{\phi}

\newcommand{\ps}{\psi}

\newcommand{\Ga}{\Gamma}
\newcommand{\De}{\Delta}

\newcommand{\La}{\Lambda}

\newcommand{\Ps}{\Psi}

\begin{document}



\title{Holographic light-front QCD}
\thanks{\begin{spacing}{0.86} \vspace{-8pt} Based on a review talk given by H. G. Dosch at the 40th anniversary of the QCD Montpellier Conference (QCD25), 30 June - 4 July 2025.  For the HLFHS Collaboration: S. J. Brodsky,  G. F. de Téramond, A.~Deur, H. G. Dosch,  Tianbo Liu, A. Paul and R. S. Sufian. \end{spacing}}

\author{Hans~G\"unter~Dosch}

\email[]{h.g.dosch@gmail.com}
\affiliation{Institut f\"ur Theoretische Physik der Universit\"at, Philosophenweg 16, D-69120 Heidelberg, Germany}

\author{Guy~F.~de~T\'eramond}
\affiliation{Laboratorio de F\'isica Te\'orica y Computacional, Universidad de Costa Rica, 11501 San Jos\'e, Costa Rica}

\author{Stanley~J.~Brodsky}
\affiliation{SLAC National Accelerator Laboratory, Stanford University, Stanford, California 94309, USA}

\date{\today}



\begin{abstract}

\noindent

This brief review presents the key components of light-front holographic quantum chromodynamics (HLFQCD). Particular attention is given to the introduction of the QCD color confinement scale within the context of a graded superconformal algebra. A concise overview of applications is provided, including  spectroscopy, form factors,  parton distributions of hadrons, and the relationship between entanglement entropy and high-energy scattering. While many other recent applications of HLFQCD exist in the literature, they are not included in this short overview of our latest work.

\end{abstract}

\maketitle

\section{General background}


Holographic light-front QCD (HLFQCD) emerged from a series of foundational articles~\cite{Brodsky:2003px, deTeramond:2005su,Brodsky:2006uqa, deTeramond:2008ht} that incorporated key insights from the Maldacena conjecture~\cite{Mal}. Formulated within the light-front quantization framework introduced by Dirac~\cite{Dirac:1949cp}, HLFQCD establishes an intuitive correspondence between a higher-dimensional gravitational theory in anti-de Sitter (AdS) space and physical QCD at its asymptotic boundary. A later development revealed that superconformal symmetry uniquely determines the form of the confinement potential, allows for the introduction of a confinement scale in the Hamiltonian, and predicts an emergent supersymmetry relating mesons and baryons~\cite{BDD14, DDBbaryon, DDBsusy}.

\vspace{8pt}

{\bf The Maldacena conjecture}  assumes that a classical theory in 5 dimensions can correspond to a quantum field theory in 4 dimensions. The classical theory considered here is a a gravitational theory with the  5th dimension, $z$, and $AdS_5$-metric $ds^2=
\frac{R^2}{z^2} \left(dx_1^2-dx_2^2-dx_3^2+dx_4^2-dz^2 \right).$
The quantized theory is a  gauge theory with massless fermions (quarks) and an infinite number of colours ($N_c \to \infty$).
 This mini-review will focus on the main considerations and results, referring the reader to the review~\cite{Brodsky:2014yha} and the original literature for details.
 
\vspace{8pt} 

{\bf Higher spin states in AdS space.} The effective action for bosons with arbitrary spin in a maximally symmetric 5-dimensional theory for bosons is given by~\cite{DDB1, dew}:
\begin{align} \label{s1}
S_{\it eff} =& \int d^{d} x \,dz \,\sqrt{g} \;  \,g^{N_1 N_1'} \cdots  g^{N_J N_J'} 
\Big(  g^{M M'} D_M \Phi^*_{N_1 \dots N_J}\, D_{M'} \Phi_{N_1 ' \dots N_J'}\nn \\
&- \mu_{\it eff}^2(z)  \, \Phi^*_{N_1 \dots N_J} \,
\Phi_{N_1 ' \dots N_J'} \Big).  
\end{align}

For higher spin states ($J > 0$), covariant derivatives must be used to preserve the AdS geometry. In non-Euclidean geometry, these derivatives ($D_N$) are rather involved. After performing a Fourier transformation on the four space-time coordinates, the resulting equation of motion can be written as follows~\cite{DDB1}:
\begin{align} \lb{Two}
\left( -\pa_z^2 +  \frac{4 L_{AdS}^2-1}{4z^2}\right)  \phi(q,z) =   q^2   \phi(q,z),
\end{align}
where the quantity $L^2_{AdS}$ varies with the spin (rank) $J $ of the AdS field in~\req{s1}, as well as depending on the on the dimensionless product of the AdS mass and curvature~\cite{DDB1}. In HLFQCD it will be identified with the LF angular momentum, as we will show below.

The resulting Hamiltonian wave equation contains no interaction term. To obtain such a term, one has to break the scaling symmetry of the AdS metric by introducing a dilaton factor $e^{\ph(z)}$ in the AdS action~\req{s1}. This gives rise to an additive interaction term  in~\req{Two}
 \begin{align}   \lb{TwoU}
 \left( -\pa_z^2 +  \frac{4 L_{AdS}^2-1}{4z^2}   +U_{AdS}(z)\right)    \phi(q,z)=   q^2   \phi(q,z),
\end{align}
where  $U_{AdS}(z) = z^\gamma  e^{-\varphi(z)/2} \, \partial_z \left(z^{-\gamma}  \partial_z e^{\varphi(z)/2} \right)$,  $\gamma = d - 1 - 2J$. The special choice $\varphi(z) = \lambda z^2$ leads to the harmonic potential $ U_{AdS}(z) = \la^2 z^2 +(2 J-1)$.

The action analogous to~\req{s1} for a spinor field $\Psi_{N_1 \cdots N_T} $ 
in  AdS space is:
\begin{align} 
S_{F \it  eff} =&\half  \int  d^d x \,dz\, \sqrt{g}  \; g^{N_1\,N_1'} \cdots g^{N_T\,N_T'} \times \nn \\
 &\left( \bar  \Psi_{N_1 \cdots N_T}  \Big( i \, \Ga^A\, e^M_A\, D_M -\mu
\Big) \Psi_{N_1' \cdots N_T'} + h.c. \right),
\end{align}
from which on can derive the equation of motion for fermions. However, in this case the introduction of a dilaton term $e^{\ph(z)}$ does not lead to an interaction potential. In order to achieve this, it is necessary to add a Yukawa term, $\bar\Ps \la_B\Ps $,  explicitly to the action~\cite{Kirsch}. The spinor field can be decomposed
into chiral components:  $ \Psi_{\nu_1
\cdots \nu_T}(q,z) = z^{2-T}\left(u^+_ {\nu_1 \cdots \nu_T}(q)
 \psi^+(q,z) +u^-_ {\nu_1 \cdots \nu_T}(q) 
\psi^-(q,z)\right)$
with
$u^\pm_ {\nu_1 \cdots \nu_T}(q)= \half(1\pm \ga_5) u_ {\nu_1
\cdots \nu_T}(q)$. This yields the following equations:
  \begin{align} 
  \left(-\pa_z^2  + \frac{4 (L_{AdS}-\half \mp \half)^2}{4z^2}+
 U^{\pm}_{AdS}(z) \right)\psi^\pm(q,z) \lb{hads}
= q^2 \psi^\pm(q,z) ,
\end{align}
\mbox{ with } $U^{\pm}_{AdS}(z)=\la_B^2 z^2 +(L_{AdS}+\half\pm \half) \la_B \lb{psib}$.
 
 \vspace{8pt}
 
{\bf Light-front quantization} is based on the quantization rules~\cite{Dirac:1949cp}  at equal light-front time $x^+=x^4+x^3 $. 
The remaining light-front (LF) coordinates are $x^- = x^4-x^3 , \, \vec x_{\perp} = (x^1, x^2)$. The coordinates of two particles --or particle clusters-- inside a hadron with total longitudinal momentum P are specified by the fractions of the total momentum $x$:  $p_{(1)} = x P; ~
 p_{(2)} = (1-x ) P$ and $\vec \ze$, the boost invariant transverse separation at equal light-front time $\vec \ze = \sqrt{ x(1- x)} \,\vec b_\perp; \vec b_\perp = (\vec x_{\perp (1)} -\vec x_{\perp (2)}).$

The simplest semiclassical bound state for a hadron is a quark-antiquark system. In the following the baryon is treated as a system of two components, namely as a quark-diquark system.  The semiclassical LF-Hamiltonian equation for two massless partons with angular momentum $L$ around the $z$-axis is~\cite{deTeramond:2008ht}
\begin{align} 
\lb{hlf} H_{LF}  \phi(\ze) =\left( -\pa_\ze^2  +
\frac{4L^2-1}{4\ze^2}+ {U(\ze)}\right)  \phi(\ze),\quad \ze=  \vert \vec \ze \vert,
\end{align}
a light-front relativistic Hamiltonian with invariant mass eigenvalues $P_\mu P^\mu = M^2$,  $H_{LF}  \phi(\ze) =  M^2 \phi(\ze)$.

\vspace{6pt}

{\bf Holographic light-front QCD (HLFQCD)} The HLFQCD framework originated in the seminal papers~\cite{Brodsky:2006uqa,deTeramond:2008ht}. These papers are based on the remarkable observation that the AdS holographic variable $z$ can be precisely identified with the LF invariant variable $\zeta$, $z = \zeta$, and that the LF Hamiltonian~\req{hlf} has the same form as the AdS wave equation~\req{TwoU}.

The main features of HLFQCD can be summarized by the following two prescriptions:\\
1) Identify the 5th holographic coordinate $z$  with the invariant separation $\zeta$ in the LF-quantized scheme.\\
2) Use the potential $U_{AdS}$ derived from the AdS  modified metric in the holographic theory as the potential $U(\ze)$ in the LF-quantized quantum field theory. 

As we shall see in Section 2, this approach yields strong results. However, at least two fundamental questions remain: \\
$\bullet$ Why do only two special procedures from a full range of possibilities lead to consistent results?
Although one can break the maximal symmetry of AdS in many ways, only a special choice of the dilaton term $e^{\la_M z^2}$ for bosons and a Yukawa term, $\bar\Ps \la_B\Ps$, lead in holographic LF QCD to phenomenologically satisfactory results for the hadron spectrum. \\
$\bullet$
Why are the relevant dimensional confinement parameters for bosons and fermions equal, $\la_M=\la_B$, despite the formally completely different schemes to introduce symmetry breaking?  This brings us to the next point.

\vspace{8pt}

{\bf Superconformal symmetry}. Holographic QCD is a child of supersymmetric string theory~\cite{Mal,Gubser,Witten:1998qj}. For some time it seemed that `the supersymmetric standard model'  was waiting around the corner, and searching for it was an important motivation to build the large hadron collider LHC. However, all efforts to find genuine superpartners of the known elementary particles have been unsuccessful thus far~\cite{PDG}.

In this work, we adopt a more modest scope by focusing solely on supersymmetric quantum mechanics~\cite{Witten:1981nf} , leaving supersymmetric field theory beyond the present discussion. It turns out that the graded algebra relevant for superconformal quantum mechanics allows for a unique breaking of conformal invariance~\cite{DDBbaryon, DDBsusy}, leading to the interaction terms discussed above, derived through two distinct approaches, and originally selected {\it ad hoc}. In practice, this can be achieved by the procedures developed by Fubini and collaborators~\cite{AFF,FR}, namely, by extending the original conformal Hamiltonian with additional terms that belong to the graded algebra of the symmetry.  We shall only present here the main points.
The conformal Hamiltonian operator of quantum mechanics in one space dimension has the same form, $H  =  -\pa_x^2 +  g/4x^2$~\cite{AFF}, as the expression following from the holographic approach~\req{Two} and~\req{hads} without the interaction potential $U_{AdS}$.

In supersymmetric quantum mechanics, the Hamilton operator $H$ is written in terms of two anti-commuting operators, the supercharges $Q$ and $Q^\dagger$, $\{Q,Q\}  = \{Q^\dagger,Q^\dagger\} = 0$, which commute with $H$,  $[Q, H]  =  [Q^\dagger,H]= 0$~\cite{Witten:1981nf, FR}. In addition to the Hamiltonian, $H$, the conformal group has two more generators corresponding to special conformal transformations, $K$, and translations, $D$. Conformal symmetry can be extended to a superconformal symmetry by adding a supercharge $S$ which anti-commutes with $Q$, $\{Q,S\}  = 0$. The supercharges $Q$ and $S$ are related to the conformal generators as follows~\cite{FR}:
\begin{align} \lb{susy1}
 \half \{Q,Q^\dagger\} = H, ~\half \{S,S^{\dagger}\}= K; \; \{Q, S^\dagger\} = \cdots . 
\end{align}
The Fubini procedure consists in a  generalization of the Hamiltonian by including also other operators of the (graded) algebra of the superconformal symmetry\,\footnote{In quantum mechanics this corresponds to a redefinition of the time variable.}. In our case we extend the Hamiltonian  $H= \half \{Q,Q\}$ to $G= \half \{R_\lambda, R^\dagger\}$ where $R_\lambda = Q + \lambda S$ commutes with the new Hamiltonian $G$.

The operators of the graded algebra  can be represented by $2 \times 2$ matrices:
\begin{align} 
Q =&\left(\!\!\!\begin{array}{cc} 0 &-\pa_x+\frac{f}{x}\\
0&0\end{array}
\!\!\!\right); \quad 
S=\left(\!\!\!\begin{array}{cc} 0 & x\\ 0&0\end{array}\!\!\!\right); \\
 H= \half\{Q,Q^\dagger\}=&
\left(\!\!\!\begin{array}{cc}
 -\pa_x^2 + \frac{4 (f + \half)^2 - 1}{4 x^2}& 0\\ 0&- \pa_x^2 + \frac{4 (f-\half)^2-1 }{4  x^2}\end{array}\!\!\!\right).
  \end{align}
 The extended Hamiltonian $G$\,\footnote{Since $Q$ has dimensions of  {\it 1/length} and K has dimensions of  {\it length}, this provides a unique  procedure~for~introducing a dimensionful quantity into the Hamiltonian.} becomes in matrix notation:\\
\begin{align}  \
G &= \half \{R, R^\dagger\} 
=\left(\!\!\!\begin{array}{cc} G_{11} & 0\\ 0&G_{22}\end{array}\!\!\!\right),  \nn
\end{align}
where
\begin{align}
\lb{g11}G_{11}&= -\pa_x^2 + \frac{4 (f + \half)^2 - 1}{4 x^2} + \la^2 \,x^2+ 2 \la\,(f-\half),\\
\lb{g22} G_{22}&= - \pa_x^2 + \frac{4 (f-\half)^2-1 }{4  x^2}+\la^2 \, x^2+ 2 \la\,(f+\half).
\end{align}

Comparing the kinetic parts of $G$, equations  (\ref{g11}, \ref{g22}), with the AdS expressions, shows that $G_{11}$ corresponds to a meson with $L=f+\half$ and $G_{22}$ corresponds to a baryon with positive chirality and $L=f-\half$.  A comparison of the interaction terms of $G$ with the modified AdS result shows also that the Fubini procedure~\cite{AFF,FR}, adopted here, leads to the same results as the two  {\it ad hoc} procedures  previously discussed, namely the the introduction of a dilaton term in the AdS action for mesons and the addition of a Yukawa term for baryons~\cite{DDBsusy, DDBbaryon}. However, the Fubini procedure turns out to be unique.

Until now, we have not explicitly taken spin into account. This can easily be done by adding to the Hamiltonian G the diagonal matrix $2 \la \cS = 2 \la s  \mbf{1}$. The value of $\it s$ is the total quark-antiquark spin for mesons or the diquark spin for baryons, i.e. $s=1$ or $s=0$ in both cases~\cite{Brodsky:2014yha}.

Finally we add a correction $\De M^2$ which takes into account the finite values of quark masses~\cite{Brodsky:2014yha}; it is  adjusted to reproduce the masses of the $\pi$ and $K$ mesons. We end up with the  final expressions:
\begin{align} 
{\mbf H}^{had}\left(\!\!\!\begin{array}{c}  \ph_{L+1}\\ \ps_{L}^+\end{array}\!\!\!\right)&=
q^2\left(\!\!\!\begin{array}{c}  \ph_{L+1}\\ \ps_{L}^+\end{array}\!\!\!\right)
 \; \mbox{ with }\lb{hfin}\\
{\mbf H}^{had} &=\left(\!\!\!\begin{array}{cc} G_{11} & 0\\ 0&G_{22}\end{array}\!\!\!\right)
+2 \la  \, \left(\!\!\!\begin{array}{cc} s_M & 0\\ 0&s_B\end{array}\!\!\!\right)+
\left(\!\!\!\begin{array}{cc} \De M^2_M & 0\\ 0& \De M^2_B \end{array}\!\!\!\right) ,\nn
\end{align}
where $G_{ii}$ are given by equations (\ref{g11}, \ref{g22})  with $f=L+\half$, $\ph_{L+1}$ and $\ps^+_L$ are the wave functions for the meson and positive chirality baryon. The orbital angular momentum of the meson is $L_M=L+1$, that of the baryon $L_B=L$, where $L$ is a non-negative integer.  The quantity $s_M$ is the intrinsic spin of the meson and $s_B$ the intrinsic spin of the diquark inside the baryon, both can have the values 0 or 1. The mass correction terms $\De M^2 _M$ and $\De M^2_B$ are due to quark masses~\cite{Brodsky:2014yha} as mentioned above.  The resulting mass spectra for mesons (M) and baryons (B) are given by  nonnegative integer values for $n$, the orbital excited states, and $L$: 
\begin{align} \lb{Mmass}
\mbox{Meson: } M_M^2 &= 4 \la (n+L_M)+ 2 \la \, s _M + \De M^2_M, \\ \lb{Bmass}
\mbox{Baryon: } 
\, M_B^2 &=4 \la(n+L_B+1) + 2  \la \, s _B  + \De M^2_B. 
\end{align}

It is important to notice that mesons with orbital angular momentum $L=0$ have no supersymmetric partners, they break the supersymmetry~\cite{DDBsusy}. In this approach,  supersymmetry exists not only between meson and baryon wave functions, but also predicts the existence of  tetraquarks as the super-partners of the baryon-wave functions with negative chirality~\cite{Brodsky:2014yha}.  We also briefly mention that the model can be extended to hadrons containing heavy quarks~\cite{DDB15, DDB16}.

\section{Applications}

{\bf Mass spectra.}  Fig.~\ref{one} shows the observed  masses of light mesons (red squares) and baryons (blue stars) as a function of angular momentum $L$. The data are compared with the prediction of the LF holographic model (dotted lines). The  predictions  depend only on one free fundamental   parameter: the confinement scale $\la$. The light quark mass corrections lead only to minor shifts in $M^2$. The agreement with the extensive data set is quite satisfactory.

\begin{figure}
\centering \hspace{-24pt}
\includegraphics[width=6.cm]{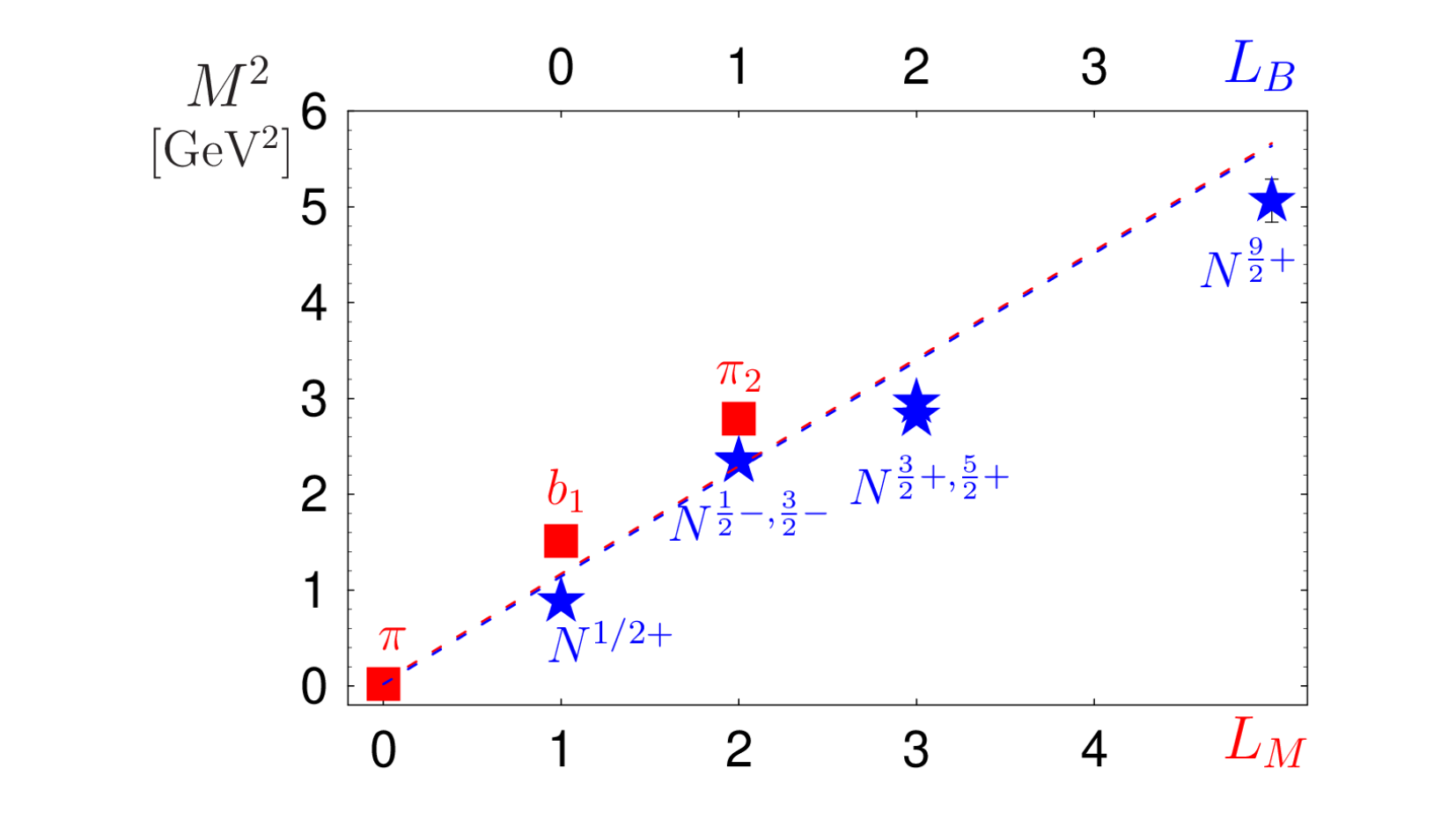} \hspace{-21pt}
\includegraphics[width=6.0cm]{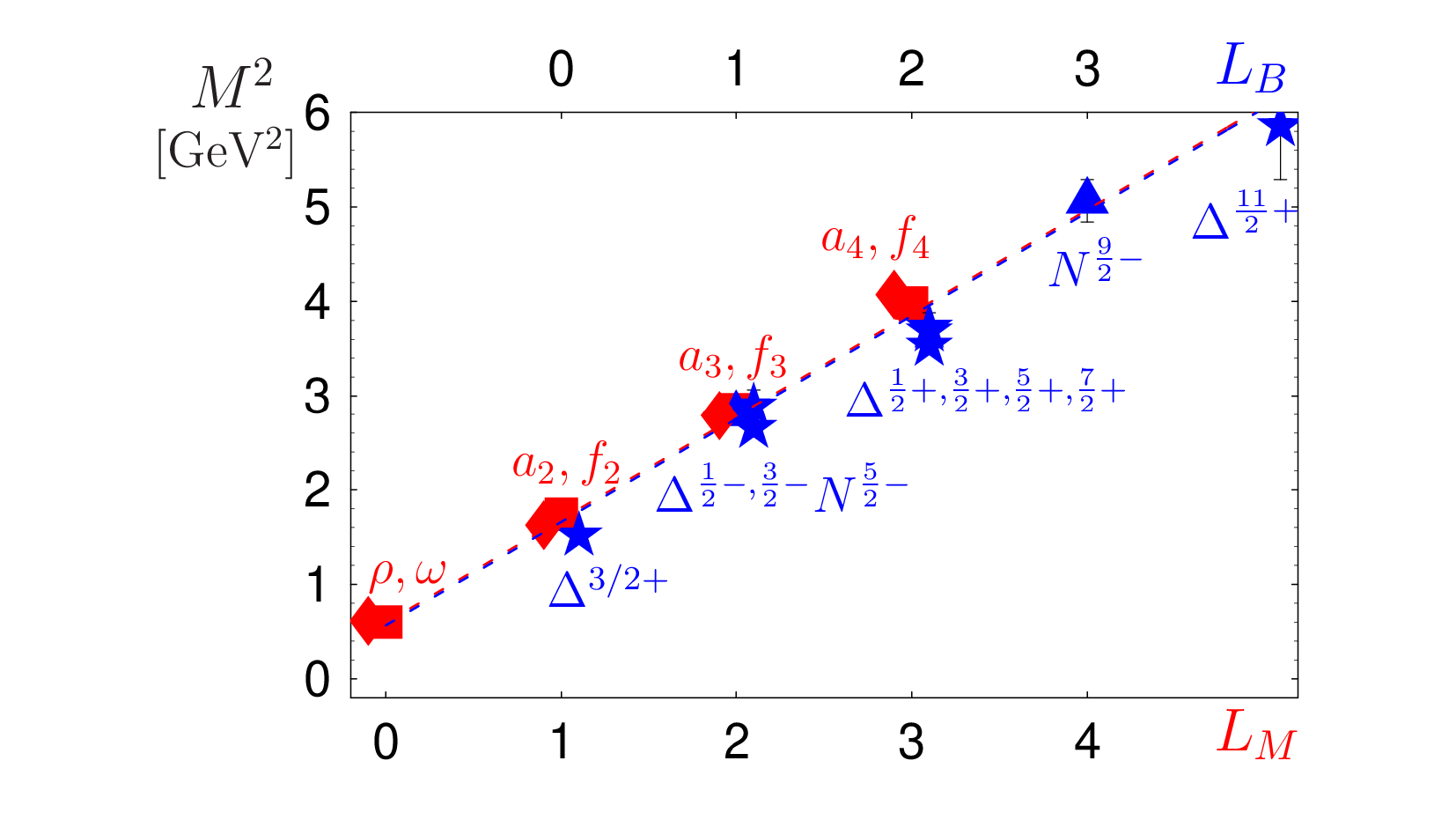}\hspace{-18pt}
\includegraphics[width=6.0cm]{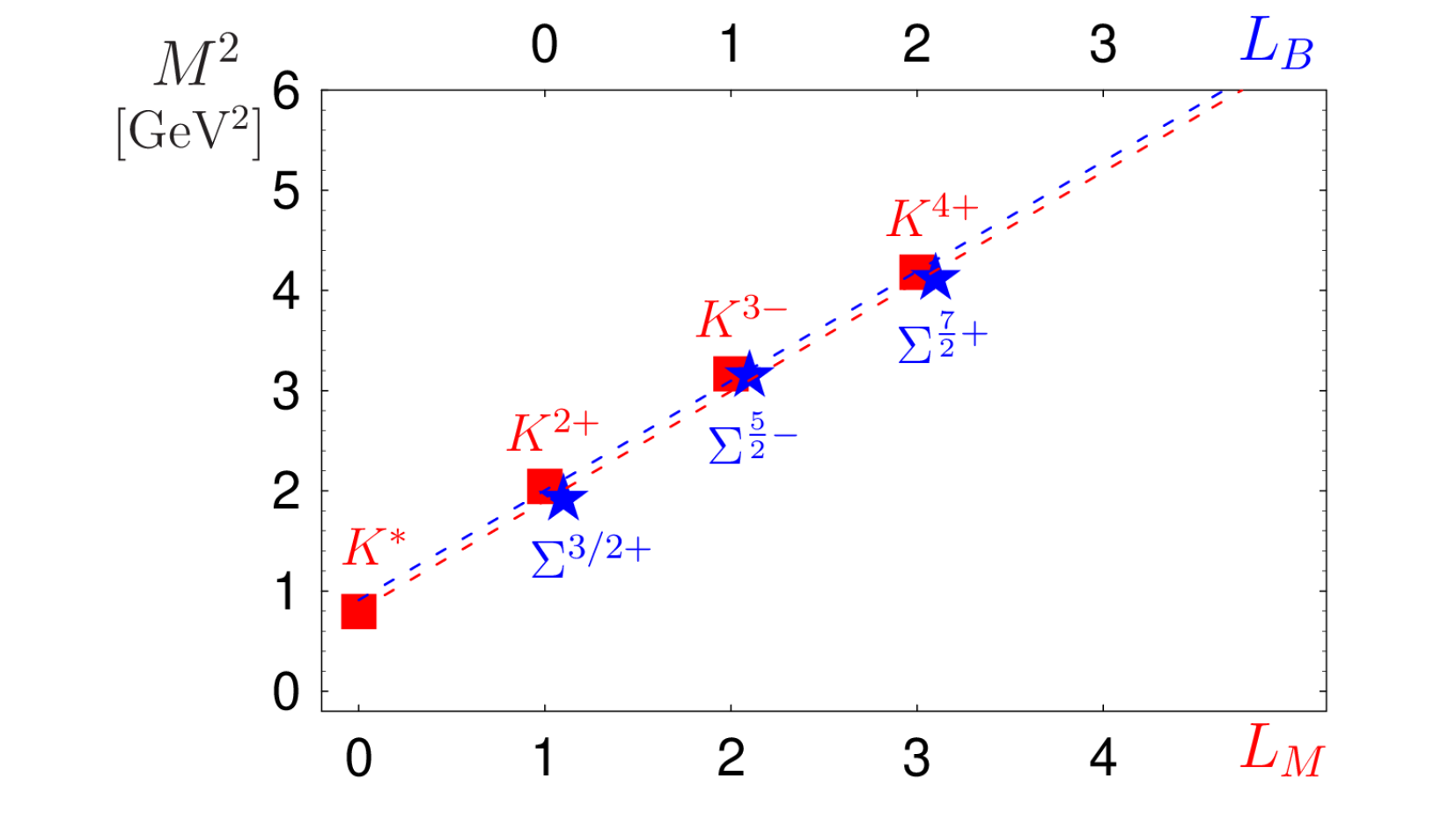}
\caption{\lb{one} Light meson and baryon trajectories as a function of the LF orbital angular momentum $L$. The dash curves are the prediction of HLFQCD  for the confinement scale $\sqrt{\la_M}= \sqrt{\la_B} = 0.52$ GeV. Finite quark masses are included by a perturbative term~\cite{Brodsky:2014yha} .}
\end{figure}

The supersymmetry between mesons with angular momentum $L+1$ and baryons with angular momentum $L$ is clearly visible, especially for $s=1$. To the best of our knowledge, the non-perturbative Fubini mechanism discussed above remains the only known explanation for this remarkable symmetry. 

\vspace{8pt}
 
{\bf Regge trajectories and the Veneziano model}. From~\req{Mmass} we can conclude that the model predicts the existence of a meson at a mass $M_M^2=q^2$, whenever the function $f_M(q^2,n) := \frac{1}{4 \lambda} q^2  + \frac{1}{2} s_M  - n - \frac{1}{4 \lambda} \De M_M^2 =  L_M+ s_M$ has the non-negative integer value $L_M + s_M = J_M$. Likewise, from \req{Bmass} we find that the function  $f_B(q^2,n) := \frac{1}{4 \lambda} q^2   + \frac{1}{2}(s_B - 1) - n - \frac{1}{4 \lambda} \De M_B^2 =  L_B+ s_B + \frac{1}{2}$ predicts the existence of a baryon at a mass $M_B^2 = q^2$, when $L_B + s_B + \frac{1}{2} = J_B$ is a non-negative half-integer number.  These relations lead immediately to an interpretation of the function $f_R(q^2,n)$ as a linear Regge trajectory~\cite{Collins:1977jy, Donnachie:2002en}. 
\begin{align}   \lb{alRn}
\al_{R,n}(q^2) =\al_{R,n}(0) +  \al'_{R,n} \, q^2 = f_R(q^2,n),   \quad \alpha_{R,n} (q^2 =M_n^2) = J,
\end{align}
where the index $R$ stands for $M$ (meson) or $B$ (baryon). The slope $\al'=1/(4 \la)$ is identical for mesons and baryons.  The intercept $\al_{R,n}(0)$
\begin{align} \lb{alintercep}
\al_{M,n}(0) = \frac{s_M}{2}  - n - \frac{\De M_M^2}{4 \la},  ~~ \al_{B,n}(0) = \frac{s_B - 1}{2}  - n - \frac{\De M_B^2}{4 \la},
\end{align}
is different for mesons and baryons. The value $n = 0$ leads to the leading Regge trajectory, and integer $n \ge 1$ to the Regge daughter trajectories. In the following we shall suppress the label $n=0$ for the leading trajectory.

\begin{figure}[ht]
\centering
\begin{minipage}[c]{.6\textwidth}
\centering
\includegraphics[width=7.6cm]{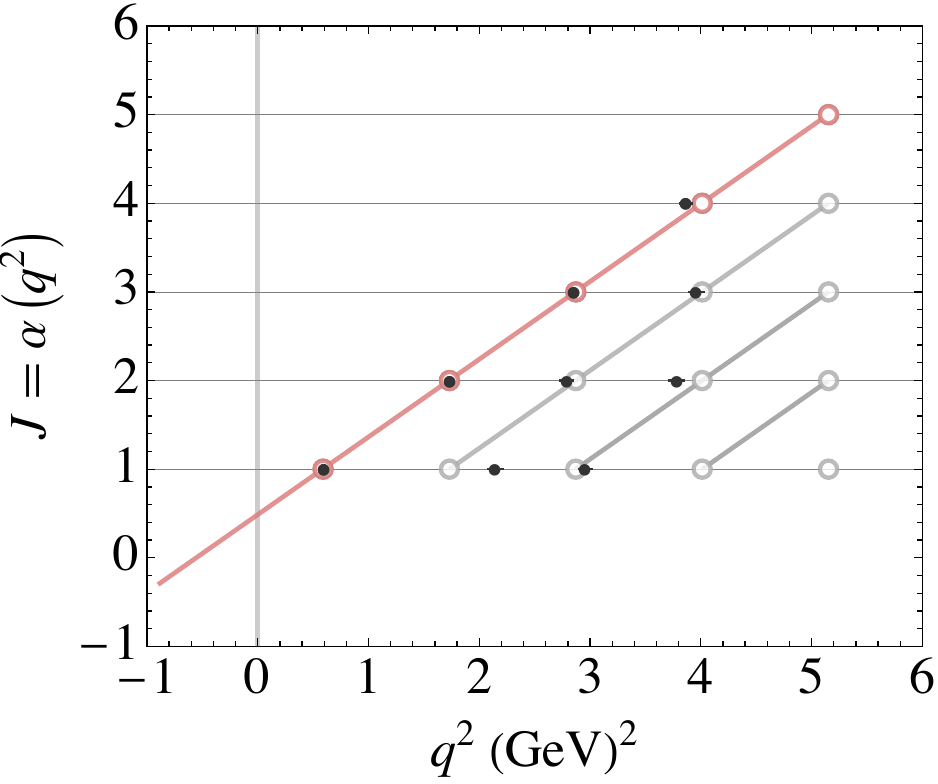} 
\end{minipage}
\begin{minipage}[c]{.35\textwidth}
\centering
\includegraphics[width=5.6cm]{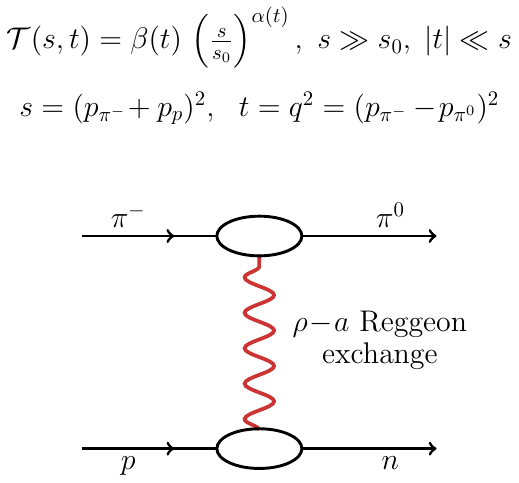}
\end{minipage}
\caption{\lb{two} The HLFQCD model predicts boson resonances that lie on linear Regge trajectories. The model predictions for light vector mesons with intrinsic quark spin $s_M=1$ (and~isospin~$I=1$) are shown as empty circles with the corresponding quantum numbers. Confirmed resonances~\cite{PDG} are represented as full circles. The leading trajectory (red) generates the the $\rho-a$ Reggeon exchange in charge exchange pion-nucleon scattering (right), with the  trajectory given by $\alpha(t)=f_M(t,0), n=0, s_M=1$. The daughter trajectories (grey), correspond to $n = 1, 2, \cdots$. These trajectories have no noticeable influence on the total cross sections at high energies due to their lower intercepts. The intercept $\alpha_{\rh}(0)$ can be extracted from the energy dependence of the cross section, $\sigma(s) \sim (s/s_0)^{\alpha_{\rho}(0)}$, and has a value close to 0.5.}
\end{figure}

The $n=0$ trajectory (red) in Fig.~\ref{two}~(left), which includes the $\rh(770), \, a_2(1320),\,\rh_3(1700)$ and $a_4(1970)$, is the leading $\rho$ trajectory, $\al_\rh(q^2)$. It dominates high-energy charge-exchange pion-nucleon scattering (Fig.~\ref{two}~(right)) because of its higher intercept. The daughter trajectories (grey) correspond to $n=1,2,\cdots$. From~\req{alRn} and~\req{alintercep}, we see that the HLFQCD model gives the following values for $\al'$ and $\alpha_\rho (0)$: $  \al'_\rh = 1/(4 \la) \simeq 0.93$ GeV$^{-2}$ and $\alpha_\rh(0) = 0.48$ for $\Delta M_M^2 = m_\pi^2 = 0.02 ~{\rm GeV}^2$, close to the chiral value 0.5 in the limit $\Delta M^2 \to 0$. For the $K^*$ trajectory one would obtain $\al_{K^*}=0.29$. Comparison of the  $t \geq 0$ behaviour with particle data~\cite{PDG}  is illustrated in Fig.~\ref{two} (left) for mesons with quark spin 1 (and isospin~1). Empty circles represent the model predictions, and full circles indicate the observed resonances along the trajectories.

The values of the $\rho$ trajectory obtained in the $t$ range including the scattering domain  $t \leq 0$, namely from -1.2 to 5 GeV$^2$ are $\al_\rh(0) = 0.5,  \, \al_\rh'= 0.9$ GeV$^{-2}$~(see \cite{Donnachie:2002en}, Fig 4.4). These values are in very good agreement with the values predicted by HLFQCD, without special recourse to Regge theory. From~\req{alintercep}, we can also compute the Regge intercept for other particle families, for example the pion, the positive parity nucleon and the $\Delta$ particle. In the chiral limit we obtain $\al_\pi(0) = 0, \, \al_N(0) = -  \,0.5, \, \al_\Delta(0) = 0$, close to the observed values.

The mass spectra  (\ref{Mmass}, \ref{Bmass}) indicates that the holographic light-front QCD has the same structure of particle masses and angular momenta  as the Veneziano model~\cite{Ven}. This model was based on general principles of $S$-matrix theory and  was very influential before the appearance of QCD. In particular, it was a source of string theory; see e.g.~\cite{DiV}.  The relationship between HLFQCD, Regge theory, and the Veneziano model reveals interesting and sometimes unexpected perspectives, as will be seen below.

\vspace{8pt}

{\bf Form factors} have been studied in holographic LF QCD from its earlier stages~\cite{Brodsky:2006uqa, BrodT08, Suf16}. Here, we only quote the basic formul\ae{} and report on other closely related applications. In AdS form factors of hadronic states are obtained as the overlap integral of the propagator of the current and the wave functions. The form factor of the conserved AdS$_5$ current is given by 
\vspace{7pt}
\begin{align}
F(t)=  R^3 \int  \frac{dz}{z^3}\,  
\Phi^*(p',z)\Phi(p,z)  A(p-p',z) , \quad t=(p'-p)^2 .
\end{align}
It is the overlap of $A(p-p',z)$, the bulk-to-boundary propagator of the conserved vector current and $\Phi_\tau(M^2,z)$, the eigenfunction of the
Hamiltonian \req{hfin}. The $z$-integration yields
\begin{align}
F^\ta(t) =  (\tau-1) B[\ta-1,1-t/(4 \la)] \lb{FFfin},
\end{align}
where $B[u,v] = \int_0^1dz\, z^{u-1}(1-z)^{v-1}$ is the Euler Beta function.

There, a serious problem becomes  apparent: The form factor obtained from the propagator of the conserved AdS-current~\req{FFfin}
leads to a zero intercept for the conserved vector current, in contrast with the intercept $\half$ from the mass eigenvalues~\req{Mmass} of the Hamiltonian~\req{hfin},  $\al_\rh(t)= \half t/(4 \la)+ \half$, in the chiral limit. To achieve consistency of the AdS  form factor results with the Hamiltonian equations, we write instead
\begin{align} \lb{FFfinB} 
F^\ta(t) =\frac{1}{\cN} B[\ta-1, 1-\al_\rh(t)]  ,
\end{align}
a simple expression which incorporates the physical intercept, as well as the quark mass terms included in the intercept~\req{alintercep}. It is important to notice that form factors of this type have been discussed previously as generalizations of the Veneziano model since 1970~\cite{Land1,Bend}. We shall use \req{FFfinB} in the following. 

\vspace{8pt}

{\bf Parton distributions of hadrons.} 
We start from the expression \req{FFfinB} for the form factor and write the integral representation of the Beta function as
\begin{align} \lb{intrep}
F_\ta(t)=\frac{1}{\cN}\int_0^1dx \,w'(x) \,w(x)^{-\al_\rh(t)}\,[1-w(x)]^{\ta-2} ,
\end{align} 
where we have substituded the integration variable  $z$ by the function  $w(x)$ with $w(0)=0, \; w(1)=1, \,  w'(x) \ge 0$~\cite{deTeramond:2018ecg}. Thus, the electromagnetic form factor~\req{intrep} can be written in terms of the longitudinal momentum fraction $x$ as $F_\tau(t) = \int_0^1 dx \, q_\tau(x) \exp[t f(x)]$ where the quark distribution function $q_\tau(x)$ and the profile function $f(x)$
\begin{align}  \lb{rho-dep}
q_\ta(x) = \frac{1}{\cN} [1-w(x)]^{\ta-2}\, w(x)^{-\al_\rh(0)}\,  w'(x), \quad  {\rm and} \quad f(x)  =  \frac{1}{4 \lambda}  \log\left(\frac{1}{w(x)}\right),
\end{align}
are both expressed in terms of the function $w(x)$.

\begin{figure} 
\centering
 \hspace{-18pt}
\includegraphics[width=7.2cm]{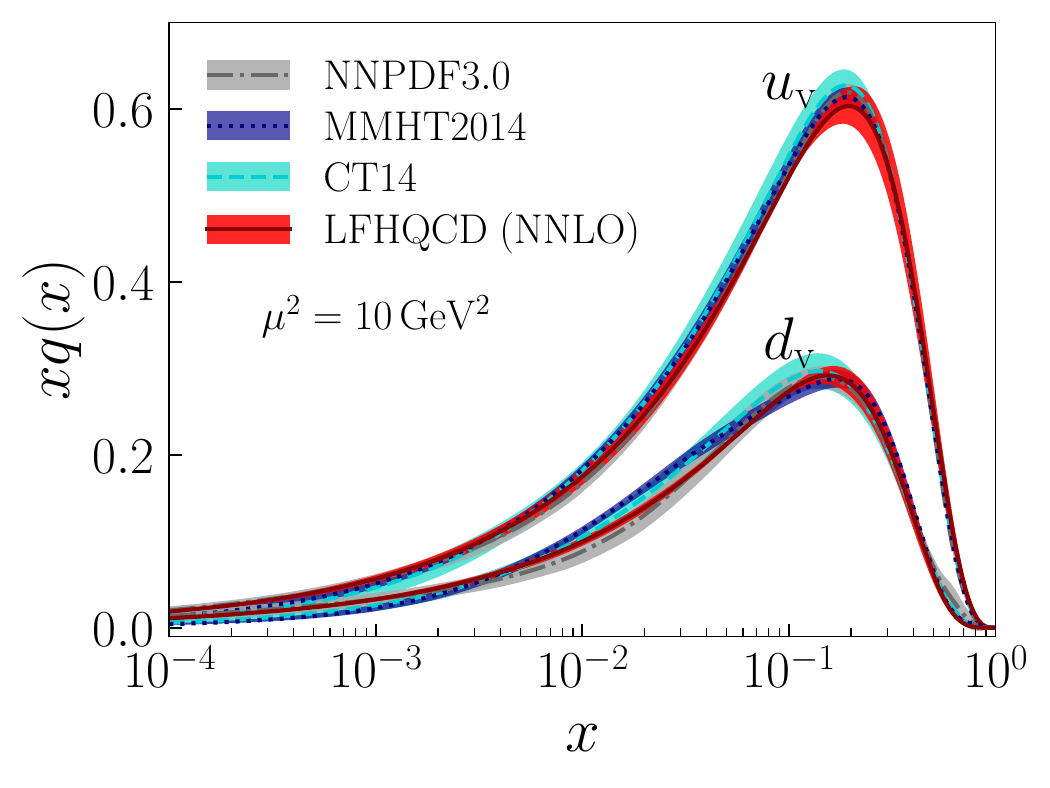} \hspace{16pt}
\includegraphics[width=7.2cm]{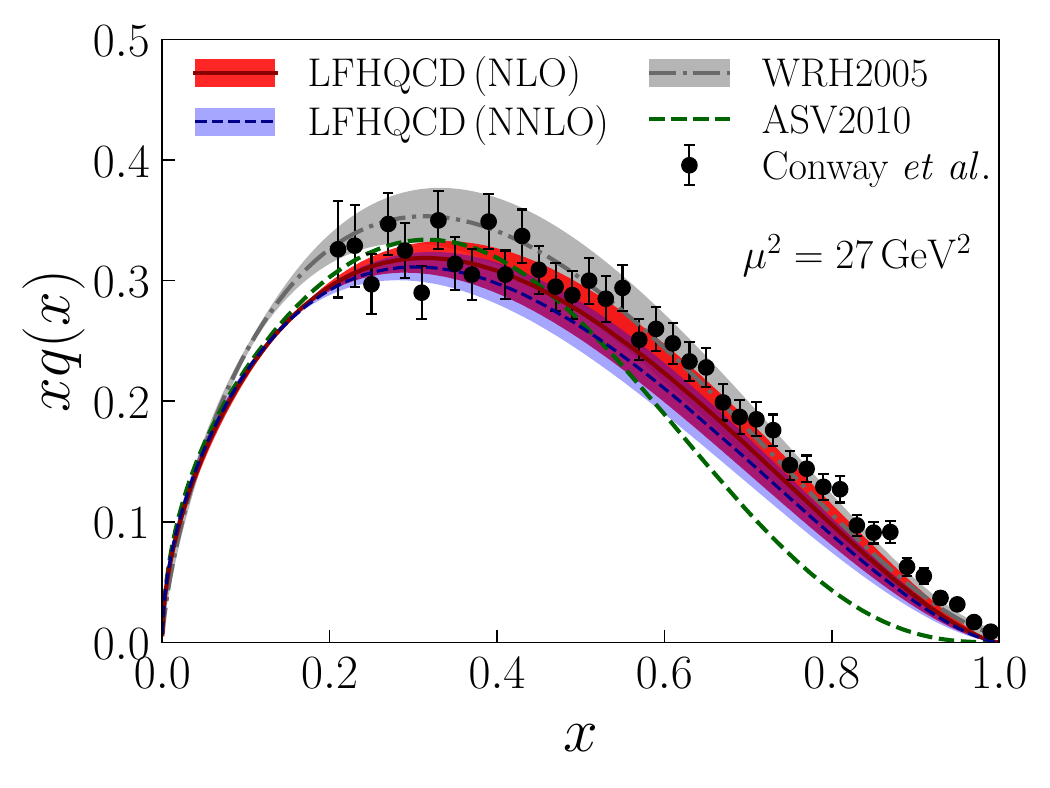}   
\includegraphics[width=7.4cm]{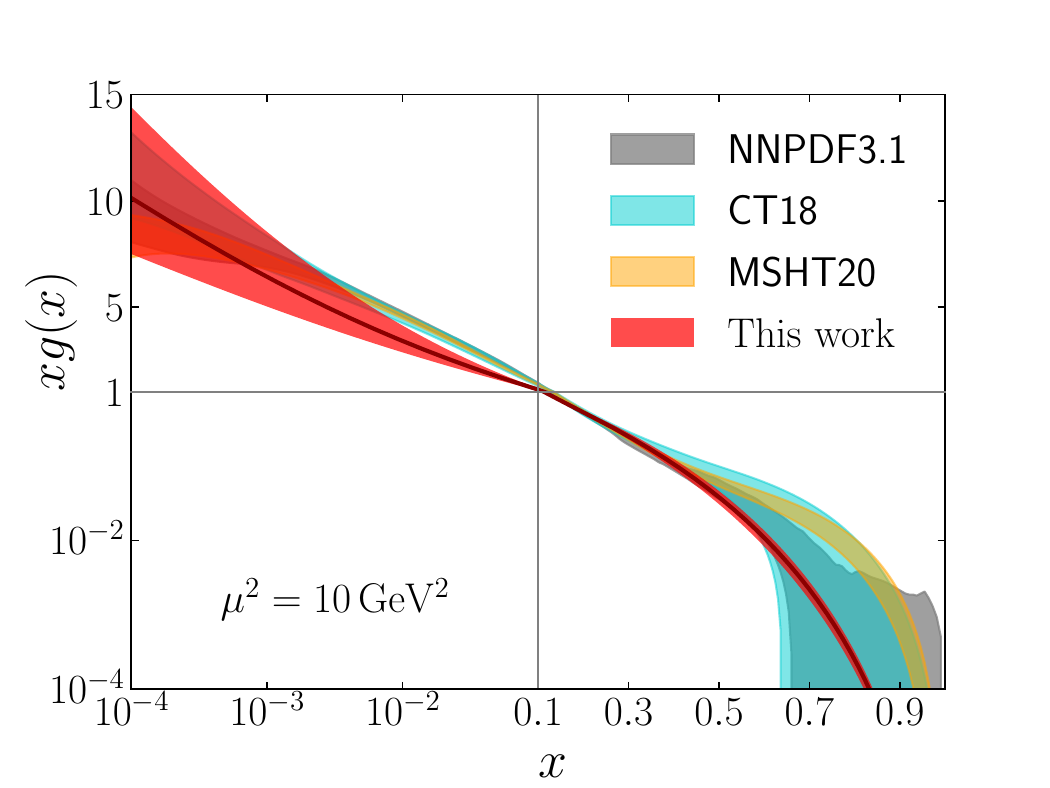} \hspace{10pt}
\includegraphics[width=7.4cm]{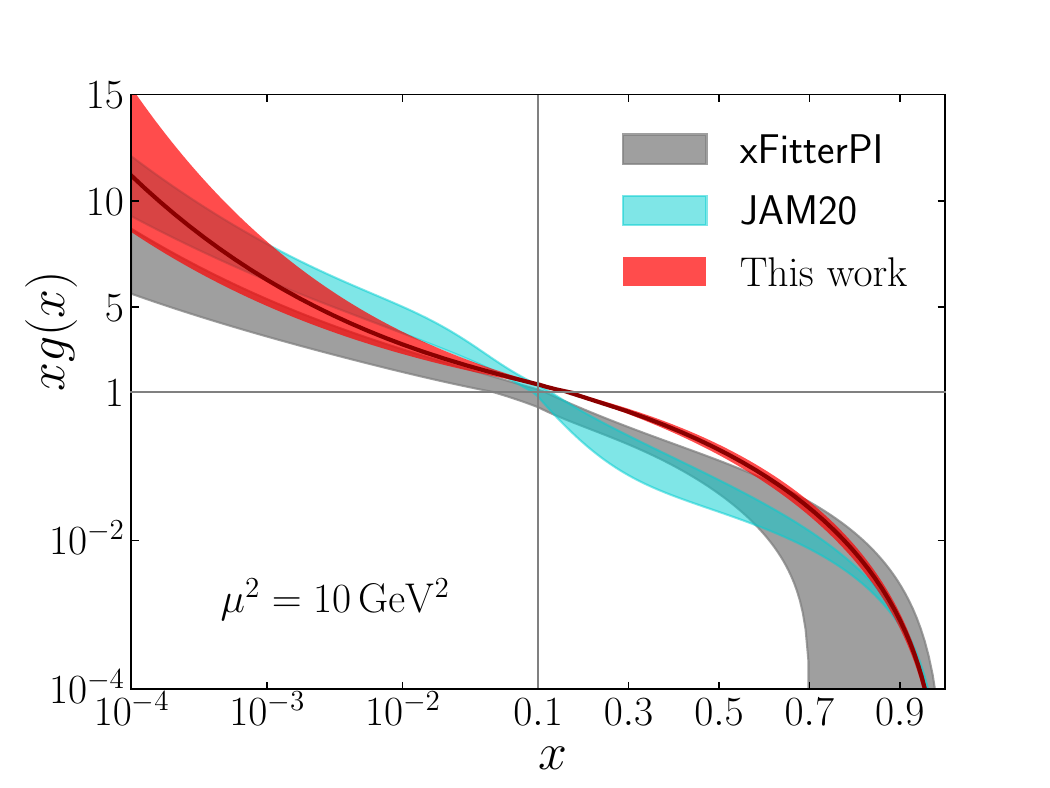} 
\caption{\lb{Three} Parton distribution functions in the proton and pion obtained in the present approach (red bands),  compared with several global fits. The results shown for the quark PDFs, $x q(x)$, are obtained by evolving the initial intrinsic distributions at the hadronic scale $\mu^2_0 \simeq 1$ GeV$^2$ to $\mu^2 = 27$ GeV$^2$ to compare with experimental results (top figures); from~\cite{deTeramond:2018ecg}. Comparison for the gluon gPDFs predictions, $x g(x)$, are evolved to $\mu^2 = 10$ GeV$^2$ (bottom  figures); from~\cite{deTeramond:2021lxc}.}
\end{figure}

The function $w(x)$ is strongly constrained by the boundary conditions at $x \to 0$ and $x = 1$; it  is a universal a universal function for all parton distributions~\cite{deTeramond:2018ecg}. With the choice $w(x)=x^{1-x} e^{-a (1-x^2)}$, and the assumption that the non-perturbative intrinsic distribution corresponds to a renormalization scale $\mu_0 \simeq 1$ GeV$^2$, we obtain the evolved results displayed in Fig.~\ref{Three} (top). Including the form factor of the axial current in the analysis one can calculate the polarized quark distributions~\cite{Liu:2019vsn}. Finally, we mention that the model, together with results from lattice gauge theory, can also determine the strange-antistrange~\cite{Sufian:2018cpj} and the charm-anticharm asymmetries in the nucleon~\cite{Sufian:2020coz}.

\vspace{8pt}

{\bf  Gluon matter distribution in hadrons}~\cite{deTeramond:2021lxc}. We can obtain the quark contribution from the vector form factor; since the latter is determined by the $\rho$-trajectory, which consists (mainly) of a quark-antiquark pair. To obtain information on the gluon distribution of a hadron,  one must consider a form factor determined by a trajectory consisting of gluons, i.e. a glueball trajectory, often identified with the Pomeron; see e.g.~\cite{Donnachie:2002en}. The lowest state on this glueball trajectory has spin 2, the same value as the graviton. Thus the corresponding tensorial form factor is usually called the gravitational form factor. The Pomeron trajectory $\al_P(t)$ has an intercept $\al_P(0)\approx 1.08$ and a slope $\al'_P \approx 0.25 $ GeV$^{-2}$.

The procedure is analogous to that used for the quark distributions. Applying the same steps as in the derivation of Eq.~\req{rho-dep}, one obtains~\cite{deTeramond:2021lxc} for the gluon density
\begin{align} x\,g_\ta(x)= \frac{1}{\cN}
[1-w(x)]^{\ta-2}\, w(x)^{1-\al_P(0)}\,  w'(x).
\lb{pom-dep} \end{align} 
The behaviour for small values of $x$ is thus determined by the Pomeron intercept. Results for the gluon distribution are shown in Fig. \ref{Three} (bottom).

\vspace{8pt}

{\bf Scale dependent Pomeron}~\cite{Dosch:2022mop}.  A  traditional description of the $Q^2$-behaviour of the electro-production processes is to assume two Pomerons, one with an intercept around 1.08 (the soft Pomeron)and the other (the hard Pomeron) with an intercept around 1.8~\cite{Donnachie:2002en}. In contrast, a single, but scale dependent Pomeron, was introduced in~\cite{Dosch:2015oha} on purely phenomenological grounds. It can explain the high energy behaviour of electroproduction with varying virtuality $Q^2$, as well as  photoproduction of heavy vector mesons. This new approach has been shown to be in accordance with the principles of Regge theory. 

\begin{figure}  
\includegraphics[width=9.6cm]{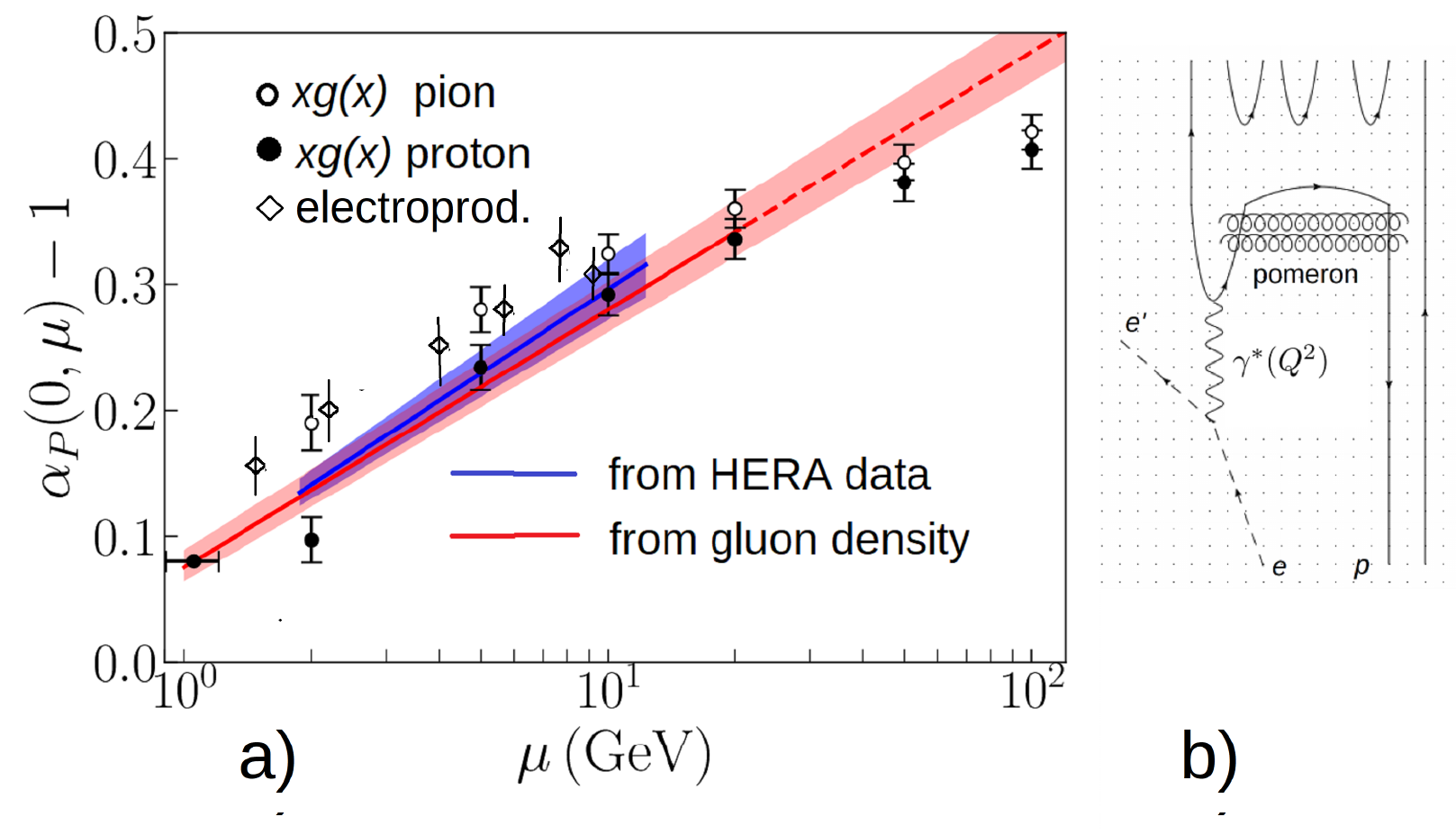}
\centering
\caption{ \lb{Four} a) The Pomeron intercept  extracted from the gluon distribution functions of the pion and proton compared with the intercept extracted from $\ga^*(Q^2)\, p$ cross section. b) 
Electroproduction off protons. The total cross section $\si(s,Q^2) \sim s^{\al_P-1}$ (structure function $F_2$)  can be measured in the electroproduction process $e+ p \to e'+X$ with $Q^2=-(p_{e'}-p_e)^2,\; s= (p_p +p_e-p_{e'})^2$.}
 \end{figure}

The close relationship between the Pomeron and the gluon density also has consequences for the Pomeron and supports the single Pomeron approach~\cite{Dosch:2015oha}.  In fact, at hadronic scales ($\mu \approx 1$  GeV) we exploited the fact that, for small values of $x$, the gluon density is determined by the Pomeron intercept; see \req{pom-dep}. Since the gluon density  depends on the scale, it is reasonable  to assume that the Pomeron intercept also depends on the virtuality scale. Fig. \ref{Four}-a  shows the behaviour of $\al_P(0,\mu^2)$ derived from the gluon distribution (circles and red line) in hadrons~\cite{Dosch:2022mop}.

The Pomeron determines the high-energy behaviour of the total cross section of virtual photon scattering off hadrons, i.e.  $\si_{\ga^*(Q^2) +p \to X}$. From the $Q^2$-dependence of the  measured cross sections one can determine the  scale dependence of the Pomeron intercept
$\al_P(0)$ as a function of the QCD scale, which is set by the photon virtuality $Q^2$. The diamonds in Fig.~\ref{Four}-a are directly extracted values~\cite{Dosch:2015oha}, the blue line is from a fit to the measured structure function $F_2$. The agreement between the two methods is satisfactory, though not unexpected. 

\vspace{8pt}

{\bf Entanglement Entropy and High-Energy scattering behaviour}~\cite{Dosch:2023bxj}. Kharzeev and Levin~\cite{Kharzeev:2017qzs} related the parton density to the entanglement entropy between the spatial region probed by deep inelastic scattering and the rest of the proton.  At small values of the Bjorken variable $x$, gluons are the dominant partons, therefore one has:  $S_{DIS}(x) = \log\left(x\,g(x)\right)$. From the relation of the intrinsic nonperturbative gluon density to the Pomeron intercept~\req{pom-dep}, $x g(x) \sim x^{1-\alpha_P(0)}$, one obtains a relation between the $x$-dependence of  entanglement entropy and  the Pomeron intercept:
\begin{align}
\frac{\pa}{\pa x} S_{DIS}(x) =  \left(\al_P(0)-1\right)\,  \log\left(\frac{1}{x}\right).
\lb{increase}\end{align}

The Bjorken variable $x$ is defined as $x= \frac{Q^2}{s+Q^2-m^2} \simeq \frac{Q^2}{s}$ at large $s$; see the caption of Fig.~\ref{Four}.  We thus have $dS_{DIS}= (\al_P(0)-1) \frac{ds}{s}$, meaning  that $(\al_P(0)-1)$ plays a similar role as the heat capacity in classical thermodynamic entropy. 

An interesting analogy exists between the increase of the entropy with squared  total energy $s$, Eq.~\req{increase}, and the rise of the total total cross section $\si_{tot}(s)$ with~$s$.  The increase of the total cross section with energy, beyond the (constant) geometrical limit --which in our case is related to  $\al_P(0)>1$ --was first shown by Heisenberg, who deduced an increase of the total hadronic cross section, such as $\log^2 s$, based on a shockwave model of colliding hadrons~\cite{Heisenberg:1952zz}. This behaviour saturates the Froissart-Martin bound~\cite{Froissart:1961ux}, which is imposed by analyticity conditions and unitarity.  In deep inelastic scattering the increase of the parton density and, correspondingly, the increase in entanglement, $S_{DIS} \sim \ln(1/x), x \ll 1$,  is a typical pair-creation effect in quantum field theory.

Finally, we note a certain analogy with the Bekenstein-Hawking entropy~\cite{Bekenstein:2008smd}, which can also be interpreted as entanglement entropy.  It can be expressed by the slogan: ``Entanglement  knows no frontiers, be they posed by black hole horizons or by confinement''.

\vspace{8pt}

{\bf  QCD effective running coupling}~\cite{Brodsky:2010ur, deTeramond:2024ikl, deTeramond:2025qlj}.  In holographic QCD, the concept of the renormalization-scale dependent running QCD coupling $\al_s(\mu^2)$ can be extended to the nonperturbative region; see e.~g.~\cite{Brodsky:2010ur}. In~\cite{deTeramond:2024ikl} the following expression for the effective coupling constant $\al_{\rm eff}$ was given:
\begin{align}\lb{gen}\
\al_{\rm eff}(Q^2) = \al_{\rm eff}(0) \exp\Big[-\int_0^{Q^2} \frac{du} {4 \la +u \log(\frac{u}{\La^2})}\Big].
\end{align}
In the non-perturbative and the perturbative regions it behaves as follows:
\begin{align}
\al_{\rm eff}(Q^2) \to \left\{
\begin{array}{lcl}
e^{-Q^2/4\la^2}, &\mbox{for } Q^2\ll 4 \la , \\
\frac{1}{\log(Q^2/\La^2)}, &\mbox{for } Q^2\gg 4 \la .
\end{array}\right.
\end{align}
For $Q^2\gg  4 \la$ this is the typical leading order perturbative behaviour. For $Q^2\ll 4 \la$, it is the non-perturbative behaviour based on the light front holographic approach~\cite{Brodsky:2010ur}.

\begin{figure}
\centering
\includegraphics[width=8.4cm]{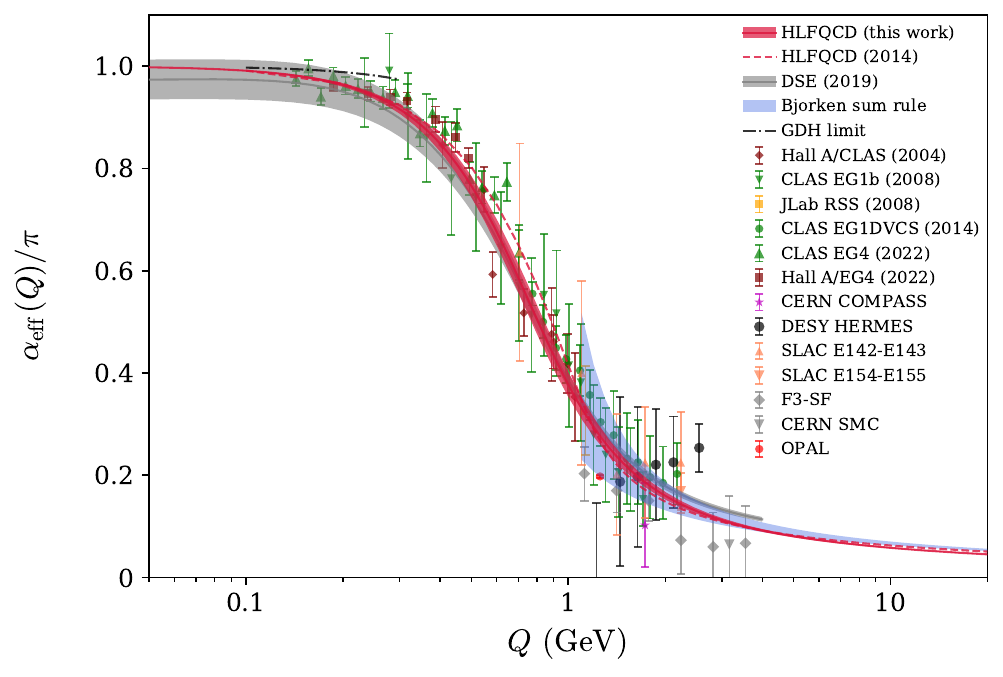}  \vspace{-4pt}
\includegraphics[width=7.9cm]{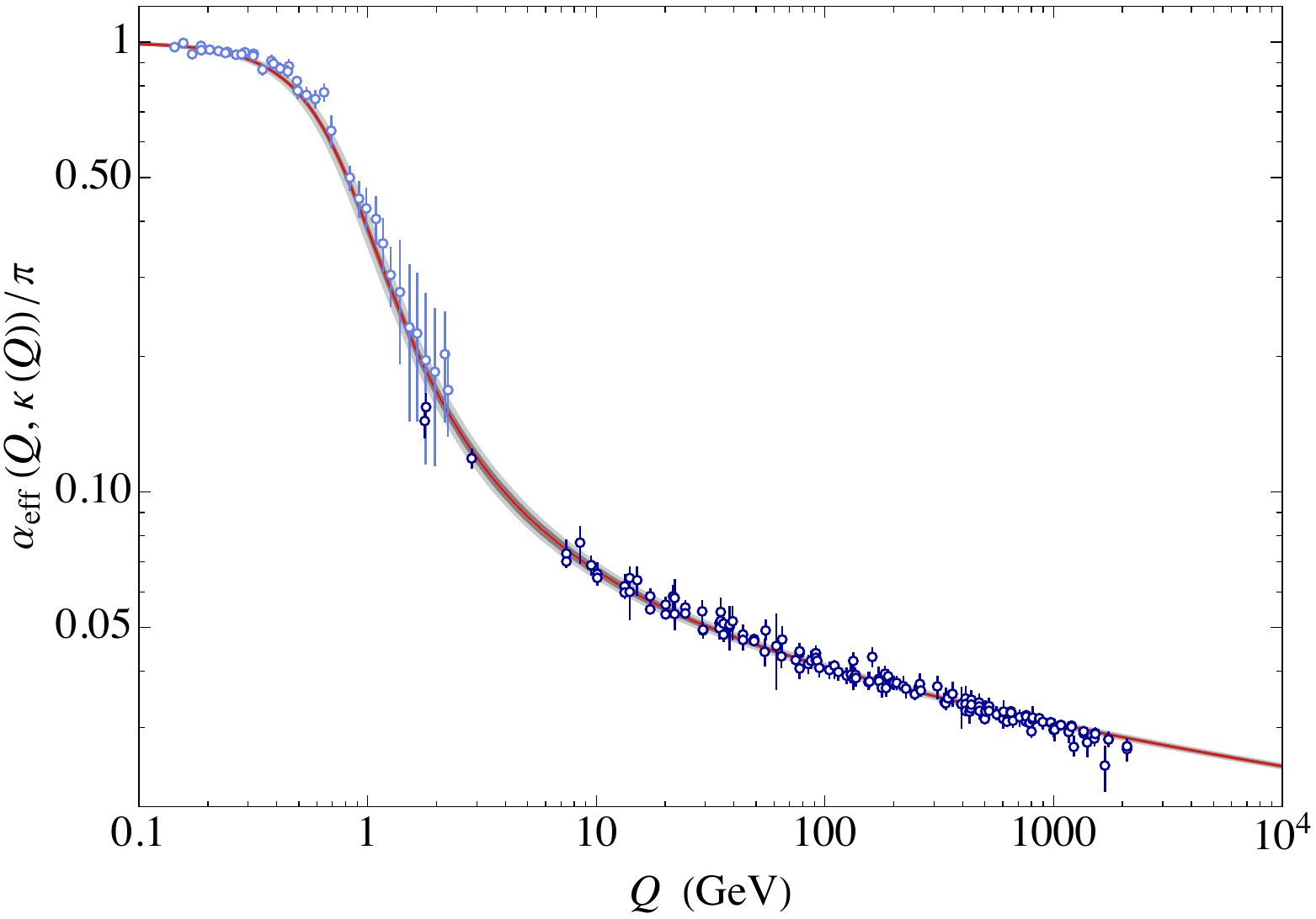}
\caption{\lb{Five} The effective coupling $\al_{\rm eff}(Q)$. According to \req{gen} with a fixed value $\la=\ka^2 = 0.286$ GeV$^2$ (left)
and with a $Q$-dependent value $\la(Q^2)= \ka^2(Q^2)$; see \cite{deTeramond:2025qlj}.}
\end{figure}

In Ref.~\cite{deTeramond:2024ikl}, it was shown that analyticity arguments allows one to  determine the perturbative constant $\La$ in terms of of the confinement scale as $\kappa$ as $\La= \frac{8}{\pi} \la$. Thus, the transition from the hadronic to the perturbative region, approximately the region $2 \leq Q \leq 5$ GeV, is also governed by the confinement scale $\la$, as can be seen from Fig.~\ref{Five} (right).  

Above this region, the influence of heavy quarks, as well as the QCD boundary conditions in the asymptotic limit $Q^2 \to \infty$, are critical elements of the model. These factors affect the $Q^2$-dependence of the running coupling, since the number of flavours enters the evolution equation. Threshold effects can be incorporated while preserving analytic properties. The model also requires the introduction of three weight coefficients, which are strongly constrained by asymptotic sum rules and reflect the onset of heavy-quark pair creation.  As a result,  one obtains the all-scales result shown in Fig.~\ref{Five} (right).

\section{Conclusions}

$\bullet$ HLFQCD is a successful and versatile model for describing the strong-coupling regime of QCD and for calculating hadron masses. It can be consistently extended into the perturbative domain, yielding nontrivial and predictive relations.

$\bullet$ This may be due to the fact that many features of the model are compatible with general features of $S$-matrix theory (see e.g., its relation to the Veneziano model~\cite{Ven}).

$\bullet$ A crucial role in the specific form of the interaction (i.e., the breaking of maximal symmetry in AdS) is played by the Fubini mechanism \cite{AFF, FR}, which is used to construct the light-front Hamiltonian for hadrons from the generators of superconformal quantum mechanics and to introduce a mass scale through the symmetries of the model.


\end{document}